\documentclass[fleqn]{article}
\usepackage{amsmath}
\usepackage{amssymb}
\usepackage{mathtools}

\usepackage{tabularx}
\usepackage[pdfversion=1.7,pdfusetitle=true,
            bookmarks=true,
            colorlinks=true,
            urlcolor=blue,
            linkcolor=blue,
            pdfborder={0 0 0}]{hyperref}

\usepackage[open,depth=6]{bookmark}
\usepackage{indentfirst}
\usepackage[backend=bibtex,natbib=true,hyperref=true,arxiv=abs]{biblatex}
\title{Distribution of elliptic twins over fixed finite fields: Numerical results}
\author{David Leon Gil}

\newcommand{\mathup}[1]{\text{\textup{#1}}}
\newcommand{\p}{\ensuremath{\mathup{p}} }

\newcommand{\BbbZ}{\mathbb{Z}}
\newcommand{\BbbE}{\mathbb{E}}
\newcommand{\BbbF}{\mathbb{F}}

\newcommand{\Ejt}{\ensuremath{\BbbE^t_j(\BbbZ_q))} }
\newcommand{\Tf}{\ensuremath{T_f(j)} }
\newcommand{\Pa}{\ensuremath{\mathup{P}_{224}} }
\newcommand{\Pb}{\ensuremath{\mathup{P}_{256}} }
\newcommand{\Pc}{\ensuremath{\mathup{P}_{384}} }
\newcommand{\Pd}{\ensuremath{\mathup{M}_{255}} }

\newcommand{\Pg}{\ensuremath{\mathup{H}_{448}} }

\newcommand{\Ej}{\ensuremath{\BbbE_j} }
\newcommand{\Ejfq}{\ensuremath{\BbbE_j(\BbbF_q)} }
\newcommand{\Ejfqt}{\ensuremath{\widetilde{\BbbE}_j(\BbbF_q)} }
\newcommand{\Ejfp}{\ensuremath{\BbbE_j(\BbbF_p)} }
\newcommand{\Ejfpt}{\ensuremath{\widetilde{\BbbE}_j(\BbbF_p)} }

\begin{document}

\maketitle

\section{Preface}

This is a preliminary note of some numerical experiments; the results
may be rather wrong.

\section{Introduction}

This paper presents the results of numerical experiments to
determine the probability, over concrete fixed finite fields,
of prime-order elliptic curves having a prime-order twist.

These curves are called ``elliptic twins'' by \autocite{ShparlinskiSutantyo},
and are useful for a variety of cryptographic applications.\footnote{\autocite{ShparlinskiSutantyo} only consider the asymptotic density of ``elliptic twins'' as
a fraction of all elliptic curves, so their results only partially
address the question of this paper. One analytic approach might be
to combine their results with the results of \autocite{GalbraithMcKee}.}

Most notable is that such curves are secure against an ``insecure
twist'' attack. This attack was introduced in 2001 by Daniel Bernstein, see
\autocite{safecurves}, who has proposed ``twist-security'' (a slightly
weaker condition) as an essential safety criterion for elliptic curves.
\autocite{curve25519}\footnote{\autocite{curve25519} autocites Burton Kaliski (\autocite{KaliskiJCryptology})
as introducing the so-called ``unsafe-twist'' attack, but I have
been unable to find any evidence either in that paper or Kaliski's thesis,
\autocite{KaliskiThesis}, that he was aware of the attack. Kaliski's
construction of an elliptic-curve-and-twist-based random number generator
does, however, require that discrete log be hard on both the curve and its
twist, as he explicitly notes.}

The most interesting result of this paper is that, for the finite
fields the NSA-generated curves are defined over, there is only an
approximately $1/100$ probability of a random prime-order curve
having a prime-order twist.

P-384 was standardized by NIST in 1999, and generated by the NSA
at some previous time.\autocite{recur} It has a prime-order twist.
\autocite{safecurves}

P-224 was standardized by NIST at the same time. It does not have
a prime-order twist. In fact, its twist has only $~58$-bit security.
\footnote{The twist of P-224 has a cofactor of $3^2 \cdot 11 \cdot 47 \cdot 3015283 \cdot 40375823 \cdot 267983539294927$. \autocite{safecurves}}

\section{Elliptic twin curves}

We follow the definitions of \autocite{ShparlinskiSutantyo}, with some minor
modifications.

Let $\Ej$ be the elliptic curve of invariant $j$, and $\Ejfq$ be its reduction
over a finite field of characteristic $p > 5, n \geq 1$ with $p$ prime. Let
and $t(\Ejfq)$ be the trace of Frobenius of that elliptic curve.

Let $\Ejfqt$ be the non-trivial quadratic twist of $\Ejfq$ over the same field.

An \emph{elliptic twin} is a pair consisting of a prime $p$, and a set of two
primes not equal to $p$ or $0$, $\lbrace l, r \rbrace$, such that

\begin{equation}
\#\Ejfp + \#\Ejfpt = l + r = 2 p + 2 - l + r
\end{equation}

\autocite{ShparlinskiSutantyo} provide evidence that elliptic twins exist over
arbitrary prime fields, but the formulae of \autocite{ShparlinskiSutantyo} do
not appear to provide precise estimates for fixed finite fields.

\section{Primes}

We consider the non-Mersenne SECP primes, standardized for the use of the
federal government in \autocite{recur}, which are, where $N \coloneqq 2^{32}$:

\begin{equation}
\begin{aligned}
    \Pa &= N^7 - N^3 + N^0                \\
    \Pb &= N^8 - N^7 + N^6 + N^3 - N^0    \\
    \Pc &= N^{12} - N^4 - N^3 + N^1 - N^0
\end{aligned}
\end{equation}

They are subset of the class of Generalized Mersennes defined by
\autocite{Solinas}.


\section{Numerical methods}

\subsection{Finding prime-order curves}

Method 1. A slightly modified version of PARI/GP was used to calculate the
traces of prime-order curves, based on code of \autocite{HamburgPARI}.
(The particular code used for this version of this paper may be
found at \autocite{junkpari}.) Point-counting was aborted early
if $\#\Ej$ was found to have a small prime factor. This computation produced
estimates for the density both of prime-order curves and elliptic twin curves
over each field. 

Method 2. For P-384, a slightly larger computation was carried out using
the same code, but set up to abort point-counting if either $\#\Ej or \#\Ejt$
had a small prime factor. The experiment 

\subsection{Results} 

Experiment 1. We calculate \Tf for each \Ej for $0 < j < 2^{20}, j \neq 1728$, then
test $\#\Ej$ and $\#\Ejt$ for primality.

For this to be a reasonable procedure, it requires the assumption that
$j$-invariant is not correlated with the probability of the curve being
an elliptic twin, even on a local scale of $2^{20}$.

Let $N_{\pi}$ be the number of prime-order curves found, and $N_{\pi'}$
be the number of elliptic twins found. Then, in this range, we have:

\begin{tabular}[l]{l|lll}
      & $N_{\pi}$ & $N_{\pi'}$ & $N_{\pi'} / N_{\pi}$ \\
  \Pa & 2790 & 31 & 1.1e-2 \\
  \Pb & 1956 & 15 & 0.8e-2 \\
  \Pc & 1131 & 20 & 1.8e-2 \\
\end{tabular}

Experiment 2. Because of the small number of elliptic twin curves found in
expriment 1, we planned to conduct the following experiment: For 1000
pseudorandomly-generated j-invariants, set $j_{i,0}$ to a j-invariant, and
increment until $j_{i,n}$ is an elliptic twin. The average of $j_{i,n}$ is
then an estimator for $1/p(pi')$.

Due to resource constraints this experiment was aborted after finding only
441 elliptic twins.

Combining these results with those of experiment 1, after bootstrapping,
give a 99\% confidence interval for $p(pi'|pi) = [0.005, 0.01]$.

\section{Future work}

In future work, we plan to extend the study to consider the more general question
of the distribution of group structure and curve exponent for reductions of
curves over fields for which their number of integral points is non-prime, and apply
similar techniques with respect to the two curves proposed for IETF use, the
nearly-Mersenne $\Pd = 2^{255}-19$ and the Hamburg-Solinas trinomial $\Pg = 2^{448}-2^{224}-1$.
\footnote{The Hamburg primes are ``Karatsuba-friendly'' and \autocite{Hamburg}
was the first to publish an algorithm that fully takes advantage of their special
form.}

(We probably won't extend this work to the Mersenne $\textup{M}_{521}$, as that
particular calculation is pestiferously large.)

\section{Conclusion}

The quantity $1 / \p(p) = N_{\pi}(\p) / N_{\pi'}(\p)$ is an estimator for
the number of trials required, when choosing a prime curve uniformly at
random in $\BbbF_q$ for that curve to be an elliptic twin.

The probability, however, that no elliptic curves in a set of $N$ curves
are elliptic twins is, of course,

\begin{equation}
1 - \lparen \prod_{0 \leq i < n} (1 - \p_i) \rparen
\end{equation}

With respect to the curves generated by the NSA for \autocite{SECP1}, and
subsequently standardized by \autocite{recur}, this calculation gives a
probability of very approximately $> 95\%$ that \emph{none} of the curves
over $\Pa, \Pb,$ and $\Pc$ would be an elliptic twin.

But the curve over P-384 \emph{is} an elliptic twin.

One might thus conclude that it is more likely than not that the NSA's 
curves were not generated by a process that samples from a uniform
distribution on prime-order curves over the chosen prime fields.\footnote{
Why, then, don't all of \Pa, \Pb, and \Pc have safe twists? Note that the
probability of that would be $\prod_{0 \leq i < n} (1 - \p_i)$, or less than
$1.5e-6$, or a roughly 1 in 630,000 chance.}

In particular, this suggests that the NSA's choice of seeds for the
``random'' prime curves were subject to additional safety criteria
not yet publicly disclosed. (Or, of course, that things with 5\%
probability aren't terribly rare events...\footnote{In partial defense of
the NSA: Suppose that it did, in fact, draw the seeds for the SECP prime
curves uniformly at random until it found prime order curves. There is no
good way of the NSA ``proving'' that it followed this procedure honestly,
even if it did. This reinforces the importance of some ``rigidity''
criterion, as per \autocite{NUMS}.})

In addition, it suggests that the fever for ``twist-security'' which
has taken grip of the cryptographic community is potentially dangerous:
These are a smallish class of elliptic curves, and there is no evidence that
-- provided an implementation is not vulnerable to a small-twist attack --
they possess either more or \emph{less} structure than a non-twist-secure
curve.

\section{Acknowledgments}

This work was inspired by Daniel Bernstein's SafeCurves website, and a
frustratingly long search for a twist-secure curve over $M_{607}$.

Many thanks to Robert Ransom for his skeptical comments, which have
helped clarify the argument of this note.

The patch to PARI/GP is derived from a patch by Michael Hamburg.

\section*{Appendix. Cofactors for SafeCurves}

This table is adapted (read stolen directly) from \autocite{safecurves}.
The rows have been sorted by the cofactor of the twist of the curve.
The curves for which twist-security was a stated security criterion
during the selection process have been omitted.

\newcolumntype{R}{>{\raggedright\arraybackslash}X}%

\begin{tabularx}{\textwidth}{llR}
\textbf{Curve}            &$h(\Ej)$ & $h(\Ejt)$ \\
\hline
\texttt{\footnotesize secp384r1     }& $\scriptstyle 1  $    & $\scriptstyle 1                                                                                        $\\
\texttt{\footnotesize secp256r1     }& $\scriptstyle 1  $    & $\scriptstyle 3 \cdot 5 \cdot 13 \cdot 179                                                             $\\
\texttt{\footnotesize secp256k1     }& $\scriptstyle 1  $    & $\scriptstyle 3^2 \cdot 13^2 \cdot 3319 \cdot 22639                                                    $\\
\texttt{\footnotesize FRP256v1      }& $\scriptstyle 1  $    & $\scriptstyle 7 \cdot 439 \cdot 11760675247 \cdot 3617872258517821                                     $\\
\texttt{\footnotesize secp224r1     }& $\scriptstyle 1  $    & $\scriptstyle 3^2 \cdot 11 \cdot 47 \cdot 3015283 \cdot 40375823 \cdot 267983539294927                 $\\
\texttt{\footnotesize brainpoolP256 }& $\scriptstyle 1  $    & $\scriptstyle 5^2 \cdot 175939 \cdot 492167257 \cdot 8062915307 \cdot 2590895598527 \cdot 4233394996199$\\
\texttt{\footnotesize brainpoolP384 }& $\scriptstyle 1  $    & $\scriptstyle 7 \cdot 11^2 \cdot 241 \cdot 5557 \cdot 125972502705620325124785968921221517             $\\
\hline
\end{tabularx}

\printbibliography

\end{document}